\documentclass[aps,pra,twocolumn,showpacs,superscriptaddress]{revtex4}

\usepackage{times,amsmath,amssymb,amsbsy,epsfig,color,graphicx,multirow}

\begin{document}

\title{Non-Gaussian entanglement swapping}

\author{F. Dell'Anno}
\affiliation{Dipartimento di Ingegneria Industriale, Universit\`a degli Studi di Salerno, Via Giovanni Paolo II, 132
84084 Fisciano (SA), Italy}
\affiliation{INFN Sezione di Napoli, Gruppo collegato di Salerno, Italy}
\affiliation{Liceo Statale P.E. Imbriani, Via Pescatori 155, I-83100 Avellino, Italy}

\author{D. Buono}
\affiliation{Dipartimento di Ingegneria Industriale, Universit\`a degli Studi di Salerno, Via Giovanni Paolo II, 132
84084 Fisciano (SA), Italy}
\affiliation{INFN Sezione di Napoli, Gruppo collegato di Salerno, Italy}

\author{G. Nocerino}
\affiliation{Trenitalia spa, DPR Campania, Ufficio di Ingegneria della Manutenzione, IMC
Campi Flegrei, Via Diocleziano 255, 80124 Napoli, Italy}

\author{S. De Siena}
\affiliation{INFN Sezione di Napoli, Gruppo collegato di Salerno, Italy}

\author{F. Illuminati}
\thanks{Corresponding author. Electronic address: filluminati@unisa.it}
\affiliation{Dipartimento di Ingegneria Industriale, Universit\`a degli Studi di Salerno, Via Giovanni Paolo II, 132
84084 Fisciano (SA), Italy}
\affiliation{INFN Sezione di Napoli, Gruppo collegato di Salerno, Italy}
\affiliation{Consiglio Nazionale delle Ricerche, Istituto di Nanotecnologia, Rome Unit, I-00195 Roma, Italy}

\date{June 9, 2016}

\begin{abstract}
We investigate the continuous-variable entanglement swapping protocol in a non-Gaussian setting, with non-Gaussian states
employed either as entangled inputs and/or as swapping resources. The quality of the swapping protocol is assessed in terms
of the teleportation fidelity achievable when using the swapped states as shared entangled resources in a teleportation protocol.
We thus introduce a two-step cascaded quantum communication scheme that includes a swapping protocol followed by a teleportation protocol.
The swapping protocol is fed by a general class of tunable non-Gaussian states, the squeezed Bell states, which,
by means of controllable free parameters, allows for a continuous morphing from Gaussian twin beams up to maximally non-Gaussian
squeezed number states. In the realistic instance, taking into account the effects of losses and imperfections, we show that
as the input two-mode squeezing increases, optimized non-Gaussian swapping resources allow for a monotonically increasing enhancement
of the fidelity compared to the corresponding Gaussian setting. This result implies that the use of non-Gaussian resources
is necessary to guarantee the success of continuous-variable entanglement swapping in the presence of decoherence.
\end{abstract}


\maketitle

\section{Introduction}

Long-distance quantum communication \cite{QCommunGisin}
is a crucial ingredient in the realization of distributed quantum information networks.
In achieving this goal, entanglement swapping plays a key role.
Such a protocol is needed, e.g., in order to realize quantum repeaters connecting distant communicating parties,
since it establishes quantum correlations between remote parties via entanglement transfer \cite{Qrepet}.
In general, the efficient teleportation of entanglement and of squeezing,
as well as entanglement purification, is a necessary requirement for the realization
of a quantum information network based on multi-step information processing \cite{QICV}.

For continuous variable (CV) systems of quantized radiation, various schemes for long-distance communication based on concatenated entanglement swapping
have been devised \cite{QKDSanders,QKDZippilli}.
A CV entanglement swapping protocol has been proposed by van Loock and Braunstein (vLB)
\cite{vanLoockBraunstein}, and demonstrated experimentally \cite{ExpSwap1,ExpSwap2,ExpSwap3}. Recently,
the transfer of discrete-variable two-mode entanglement has been experimentally demonstrated
in a hybrid framework,  by exploiting CV resources and operations \cite{HybridSwap}.
The entanglement swapping protocol requires the exploitation of suitably entangled CV resources.
The simplest available ones are two-mode Gaussian states, and a detailed analysis of the optimal Gaussian entanglement swapping
has been carried out in Ref.~\cite{OptimalGaussSwapp}.
On the other hand, it has been shown that selected classes of non-Gaussian CV states can,
in principle, be powerful for the efficient implementation of quantum information and metrology tasks
\cite{CVTelepNoi,RealCVTelepNoi,CVTelepNoisyNoi,CVSqueezTelepNoi,KimBS,DodonovDisplnumb,Cerf,Opatrny,Cochrane,Olivares,KitagawaPhotsub,YangLi,QEstimNoi,QCMenicucci,NGBartley,NonGaussTelepChina}.

Due to their high degree of non-classicality, non-Gaussian resources may offer
a better performance with respect to their Gaussian counterparts.
Indeed, among all CV states with the same fixed first and second statistical moments,
Gaussian states are the ones that \emph{minimize} various nonclassical properties \cite{ExtremalGaussian,Genoni}. In addition,
distilling Gaussian states using only Gaussian operations is impossible \cite{Scheel}, and insuperable limitations to the transport of logical quantum information arise when using Gaussian cluster states (even when arbitrary non-Gaussian local measurements are allowed), implying the need for non-Gaussian resources in measurement-based quantum computing \cite{EisertLimitations}.
Many theoretical and experimental efforts have been devoted to the engineering of non-Gaussian states of the radiation field
(for a review on quantum state engineering, see e.g. \cite{PhysRep}).

Various theoretical methods for the generation of non-Gaussian states have been proposed \cite{CxKerrKorolkova,NonlinBogoNoi,AgarTara,DeGauss1,DeGauss2,DeGauss3,DeGauss4,DeGauss5,DeGauss6,DeGauss7,DeGauss8,DeGauss9},
several successful experimental realizations have been reported \cite{ZavattaScience,ExpdeGauss1,ExpdeGauss2,Solimeno1,Grangier,BelliniProbing,GrangierCats,Solimeno2,ExpEtesse,ExpHuang},
and different criteria have been devised for the characterization and/or quantification
of non-classicality \cite{nonclassty1,nonclassty2,nonclassty3,nonclassty4}
and of non-Gaussianity \cite{nongaussty1,nongaussty2,nongaussty3,nongaussty4,nongaussty5,nongaussty6}.

The squeezed Bell ($SB$) states introduced and investigated in Refs.~\cite{CVTelepNoi,RealCVTelepNoi,CVTelepNoisyNoi}
form a particularly interesting class of non-Gaussian entangled resources, characterized by free parameters that can be tuned to obtain known Gaussian and non-Gaussian states, including, among others, twin beams, photon-added and photon-subtracted squeezed states, and squeezed number states.
The free parameters in the class of BS states allow for significant degrees of optimization in implementing quantum protocols.
For instance, optimized $SB$ resources allow for a teleportation fidelity
higher than that associated with resources such as Gaussian twin beams or non-Gaussian photon-subtracted squeezed states
(these last being currently the best experimentally generated resources) for a large variety of teleported input states,
including coherent, squeezed, and number states \cite{CVTelepNoi,RealCVTelepNoi,CVTelepNoisyNoi}.
Simple schemes for the generation of $SB$ states have been recently proposed, based on the exploitation of independent twin beams and suitable conditional coincidence measurements \cite{SqueezBellEngineer}.

In the present work we investigate the performance of $SB$ non-Gaussian entangled resources in the implementation
of the CV entanglement swapping protocol. In our analysis the $SB$ states are exploited
as two-mode entangled input states and/or two-mode entangled resources.

Given two Bosonic field modes $h, k$, the pure, two-mode $SB$ states $|\psi_{hk} \rangle_{SB}$ are defined as:
\begin{equation}
|\psi_{hk}\rangle_{SB} =
S_{hk}(\zeta_{hk}) \{\cos\delta_{hk} |0,0 \rangle_{hk} + e^{i \theta_{hk}} \sin\delta_{hk} |1,1 \rangle_{hk} \} \, ,
\label{squeezBell}
\end{equation}
where $S_{hk}(\zeta_{hk}) = e^{ -\zeta_{hk} a_{h}^{\dag}a_{k}^{\dag} + \zeta_{hk}^{*}
a_{h}a_{k}}$ is the two-mode squeezing operator, $\zeta_{hk}=r_{hk} e^{i \phi_{hk}}$
is the squeezing complex parameter, and
$|n \, , n \rangle_{hk}  \equiv |n \rangle_{h} \otimes |n \rangle_{k}$
is a two-mode Fock state with $n$ photons in each mode, associated to modes $h$ and $k$.
The free tunable parameters $\delta_{hk}, \theta_{hk}$ control the degree of non-Gaussianity,
allowing to span a great variety of states, from Gaussian twin beams ($TB$) to squeezed number states ($SN$),
through intermediate non-Gaussian states which include, among others, the photon-added ($PA$) and the
photon-subtracted ($PS$) squeezed states.

In Tab.~\ref{tableAcro} we provide a list of the most relevant states, denoted by acronyms,
representing special realizations of the general class of two-mode $SB$ states
(\ref{squeezBell}), together with the corresponding values of the free parameters that realize them.

\begin{center}
\begin{table}[h]
\centering
\begin{tabular}{|c|c|c|}
  \hline
         Squeezed Bell states & $SB$ & arbitrary $\delta_{hk}$, $\theta_{hk}$ \\
  \hline\hline Twin Beams & $TB$ &  $\delta_{hk}=0$,  $\theta_{hk}=0$ \\
\hline  Photon Subtracted squeezed states & $PS$ & \begin{tabular}{c}
                                                   $\cos\delta_{hk}=\frac{\cosh r_{hk}}{\sqrt{\cosh 2r_{hk}}}$ \\
                                                   $\theta_{hk}=\phi_{hk}$
                                                 \end{tabular}  \\
\hline  Photon Added squeezed states & $PA$ & \begin{tabular}{c}
                                                   $\cos\delta_{hk}=\frac{\sinh r_{hk}}{\sqrt{\cosh 2r_{hk}}}$ \\
                                                   $\theta_{hk}=\phi_{hk}$
                                                 \end{tabular} \\
\hline  Squeezed Number states & $SN$ & $\delta_{hk}=\frac{\pi}{2}$, $\theta_{hk}=0$ \\
  \hline
    \end{tabular}
    \caption{List of particular states belonging to the general class of squeezed Bell states.}
    \label{tableAcro}
    \end{table}
    \end{center}

When used as entangled resources, optimized $SB$ states allow for improved performance
of the CV quantum teleportation protocol compared to the corresponding Gaussian $TB$ states with the same covariance matrix,
especially at low and intermediate levels of two-mode squeezing $r_{h, k}$~\cite{CVTelepNoi,RealCVTelepNoi,CVTelepNoisyNoi}.
Of course, in the ideal case, the improvement becomes more and more marginal as squeezing increases, and vanishes asymptotically
in the limit of infinite squeezing, as both $SB$ and $TB$ states converge to the Einstein-Podolski-Rosen (EPR) maximally entangled state.

Within the family of $SB$ states, more modest improvements with respect to $TB$ resources can also be obtained with $PS$ resources.
On the other hand, $PA$ and $SN$ resources, although both highly entangled and non-Gaussian,
do not allow for efficient teleportation.

When using non-Gaussian resources,
optimization of the teleportation performance does not simply correspond
to a maximum amount of shared entanglement, as one would naively expect and as is indeed the case when using Gaussian resources~\cite{AdessoIlluminati2005}.
Rather, optimization requires a fine interplay among the optimization of three quantities:
1) entanglement; 2) degree of non-Gaussianity (as suitably quantified by proper entropic or geometric measures at fixed covariance matrix); 3) squeezed-vacuum-affinity (defined as the overlap between the specific two-mode state considered and the two-mode squeezed vacuum)~\cite{CVTelepNoi}.
Moreover, $SB$ states allow for different optimization procedures depending on the quantity to be teleported~\cite{CVSqueezTelepNoi}. Therefore, the level of performance of non-Gaussian resources depends on the task to be accomplished, i.e. the target of the protocol these resources are used for.

We investigate the performance of the CV vLB entanglement swapping protocol when non-Gaussian $SB$ states
are used either as entangled inputs and/or as entangled resources.
Quantitatively, the level of performance is assessed by introducing a cascade scheme: we will consider the (ideal) teleportation fidelity of an input coherent state when the non-Gaussian swapped entangled states (output of the swapping protocol) are exploited as entangled resources of the CV BKV teleportation protocol.

In the first two steps we analyze the ideal and the realistic swapping protocol using as input states and/or resources \emph{generic} $SB$ states, that is, with tuning parameters left free and not optimized.
Next, we compute (in the ideal instance) the corresponding fidelity in the teleportation of a coherent state
when the (generic) non-Gaussian swapped states are exploited as entangled resources, and we maximize it over the set of free parameters;
these include the tunable parameters of the $SB$ states and the experimental gains \cite{TelepGainBowen}.
In this way, we identify, within the family of $SB$ states, the particular state that ensures the best teleportation performance when
used as a resource \emph{after} the swapping procedure has been implemented. Finally, adopting the same criterion,
we compare its performance with that of other swapped Gaussian and non-Gaussian resources.

We study the CV swapping protocol with non-Gaussian inputs and/or resources by extending to this case the formalism of the characteristic function that
has been introduced in Ref. \cite{MarianCVTelep}. Here we follow the same technical procedure as in Ref.~\cite{RealCVTelepNoi}.
The phase-space description proves to be particularly appropriate in the instance of CV non-Gaussian states,
as the corresponding mathematical machinery turns out to be computationally effective.

The paper is organized as follows.
In section \ref{secSwap}, we briefly review the entanglement swapping protocol in the formalism of the characteristic function, and we introduce the criterion by which one can assess the performance of the non-Gaussian entangled resources.
In section \ref{secResults} we present and discuss our main results, including the analysis, in comparative terms, of the performance of the entanglement swapping protocol implemented with non-Gaussian resources, both in the ideal case and in realistic instances.

\section{CV entanglement swapping protocol}
\label{secSwap}

\subsection{Characteristic functions}

Here we briefly review the two-mode CV entanglement swapping protocol. The task is the transfer of two-mode entanglement between a pair of field modes initially prepared in a entangled state, say modes $1$ and $2$, to a pair of modes initially disentangled, say modes $1$ and $4$.
Such a task is accomplished by exploiting another initially prepared entangled state of modes $4$ and, say, $3$, a Bell measurement,
and finally a pair of local unitary transformations.
Three users (Alice, Bob, and Charlie) are involved in the protocol.
Initially,  Alice shares the two-mode input entangled state of modes $1$ and $2$ with Charlie,
while Bob shares with Charlie the two-mode entangled resource of modes $3$ and $4$.

\begin{figure}[b]
  \centering
  \includegraphics[width=7.5cm]{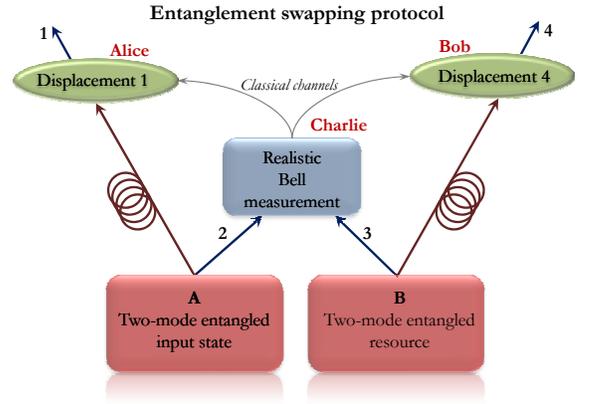}
  \caption{(Color online)
  Pictorial representation of the non-ideal CV entanglement swapping protocol.
  Initially, two users, say Alice and Bob, share one entangled state each with a third party, say Charlie.
  Alice shares with Charlie the input two-mode entangled state of modes $1$ and $2$, and
  Bob shares with Charlie the two-mode entangled resource of modes $3$ and $4$.
  In the first step, at Charlie's location, the mode $2$ of the input two-mode entangled state is mixed
  with the mode $3$ of the entangled resource.
  The ensuing state is then subject to a realistic Bell measurement (imperfect photodetection).
  The result of the measurement is communicated by Charlie to Alice and Bob through classical channels.
  The modes $1$ and $4$ propagate towards the corresponding locations through noisy channels, e.g. optical fibers.
  In the second step, two local unitary transformations, determined by the previous measurement,
  are applied by Alice and Bob to mode $1$ and $4$, respectively.
  The ensuing output state of modes $1$ and $4$ is the final swapped (entangled) state.
  Such a state is shared by the two final users Alice and Bob.}
  \label{FigSwapping}
\end{figure}

A schematic picture and a brief description of the CV swapping protocol are provided in Fig.~\ref{FigSwapping}.
In this scheme, mode $2$ of the two-mode input entangled state is mixed to mode $3$ of the entangled resource
at a balanced beam splitter.
A Bell measurement, realized by homodyne detections, is performed on the mode
resulting by the mixing of modes $2$ and $3$.
In order to model a non-ideal measurement, or equivalently to simulate the inefficiencies of the photodetectors,
a further fictitious beam splitter is placed in front of each ideal detector \cite{LeonhardtRealHomoMeasur}.
After the realistic Bell measurement, the result is transmitted through classical channels
to the locations of modes $1$ and $4$.

It is assumed that both the input state and the resource are produced close to the Charlie's location (Bell measurement),
and far from Alice's and Bob's locations (remote users), so that the modes are spatially separated.
Therefore, it can be supposed that the modes $2$ and $3$ are not affected by the decoherence due to propagation;
on the contrary, the modes $1$ and $4$ propagate through noisy channels, e.g. optical fibers,
towards Alice's and Bob's locations, respectively.
According to the result of the Bell measurement, at these locations local
unitary displacements are performed on mode $1$ of the input state and on mode $4$ of the resource.
The resulting two-mode swapped (entangled) state of modes $1$ and $4$ is the output state of the protocol.
The protocol can be described in the characteristic function formalism, as detailed in the following
(see also appendix \ref{appsecCharFuncSwapp} for further mathematical details).

Let $\rho_{0}= \rho_{12}^{A} \otimes \rho_{34}^{B}$ be the global input bi-separable four-mode state.
In phase space with quadrature field variables $(x_i,p_i) \,, i=1,\ldots 4$, such a state is described
by the characteristic function $\chi_{0}(x_1,p_1;x_2,p_2;x_3,p_3;x_4,p_4)$:
\begin{eqnarray}
&&\chi_{0}(x_1,p_1;x_2,p_2;x_3,p_3;x_4,p_4)= \nonumber \\
&& \chi_{12}(x_1,p_1;x_2,p_2) \; \chi_{34}(x_3,p_3;x_4,p_4) \,,
\label{chi0xp}
\end{eqnarray}
where $\chi_{12}(x_1,p_1;x_2,p_2)$ and $\chi_{34}(x_3,p_3;x_4,p_4)$
correspond to the characteristic functions of the two-mode entangled states
$\rho_{12}^{A}$ and $\rho_{34}^{B}$, respectively.
The output characteristic function $\chi_{out}(x_1,p_1;x_4,p_4)$ associated with the two-mode entangled output state
of the swapping protocol is given by:
\begin{eqnarray}
&&\hspace{-.5cm}\chi_{out}^{(swapp)}(x_1,p_1;x_4,p_4) =
\nonumber \\
&&\hspace{-.5cm} \chi _{12}\big( e^{-\frac{\tau_1}{2}}x_{1},e^{-\frac{\tau_1}{2}}p_{1}; T_2 (g_1 x_1 + g_4 x_4), T_3 (-g_1 p_1 + g_4 p_4) \big) \nonumber \\
&&\hspace{-.5cm} \chi _{34}\big( T_2 (g_1 x_1 + g_4 x_4), -T_3 (-g_1 p_1 + g_4 p_4); e^{-\frac{\tau_4}{2}}x_{4} , e^{-\frac{\tau_4}{2}}p_{4} \big) \nonumber \\
&&\hspace{-.5cm}  e^{-\frac{1}{2} (1-e^{-\tau_1})\left(\frac{1}{2}+n_{th,1}\right)(x_{1}^{2}+p_{1}^{2})-\frac{1}{2} (1-e^{-\tau_4})\left(\frac{1}{2}+n_{th,4}\right)(x_{4}^{2}+p_{4}^{2})}  \nonumber \\
&&\hspace{-.5cm} e^{-\frac{R_2^2}{2}(g_1 x_1 + g_4 x_4)^2-\frac{R_3^2}{2}(-g_1 p_1 + g_4 p_4)^2} \, ,
\label{chiout}
\end{eqnarray}
where $T_i$ $(R_i)$, with $i=2,3$, are the transmissivities (reflectivities) associated with the fictitious beam splitters
that model the inefficiencies of the homodyne detections;
$g_i$ ($i=1,4$) are the gains associated with the unitary displacements;
$\Upsilon_i$ and $n_{th,i}$ ($i=1,4$) are, respectively, the damping factors and the average numbers of thermal photons
associated with the noisy channels. Finally, $\tau_i$ denotes the dimensionless time $\tau_i= \Upsilon_i t$.
For a better understanding we list in Tab.~\ref{tableExpPara} all the parameters
appearing in expression~(\ref{chiout}) and associated with the experimental apparatus.
Such parameters are assumed to be fixed constants; indeed, we assume a complete knowledge
of the experimental apparatus' components.

\begin{center}
\begin{table}[h]
\begin{tabular}{|c|c|}
  \hline
         $g_i$ ,  $i=1,4$ & gains associated with unitary displacements \\
  \hline  $T_i$ $(R_i)$ , $i=2,3$ & transmissivities (reflectivities) at beam splitters \\
  \hline $\Upsilon_i$ , $i=1,4$ & channel damping factors \\
\hline  $n_{th,i}$ , $i=1,4$ & average numbers of thermal photons \\
\hline  $\tau_i \equiv \Upsilon_i t$ , $i=1,4$ & dimensionless times  \\
  \hline
    \end{tabular}
    \caption{Parameters characterizing the experimental apparatus.}
    \label{tableExpPara}
    \end{table}
    \end{center}

In the instance of an ideal protocol $(R_i=0\,,T_i \equiv 1-R_i = 1\,,\tau_i=0)$
and for $g_1=0$, $g_4=1$, Eq.~(\ref{chiout}) reduces to:
\begin{eqnarray}
&&\chi_{out}^{(swapp)}(x_1,p_1;x_4,p_4) =
\nonumber \\
&& \chi _{12}\left( x_{1},p_{1}; x_{4} , p_{4} \right) \;\; \chi _{34}\left(  x_{4}, - p_{4}; x_{4} , p_{4} \right)
 \,.
\label{chioutideal}
\end{eqnarray}
This last formula offers a clear interpretation of the task of the swapping protocol.
Assuming the entangled resource to be a twin beam with squeezing parameter $r_{34}$,
in the limit of large $r_{34}$ (perfect EPR resource) the function $\chi _{34}\left(  x_{4}, - p_{4}; x_{4} , p_{4} \right) \rightarrow 1$;
correspondingly, the output characteristic function $\chi_{out}$
coincides with $\chi_{12}$, with the complete swapping of mode $2$ with mode $4$.

\subsection{Swapping efficiency}

In order to assess the efficiency of the swapping protocol applied to input Gaussian or non-Gaussian entanglement and implemented with Gaussian or non-Gaussian resources, we proceed as follows. We study the performance of the output states produced by the swapping protocol (two-mode swapped states),
as they are used as entangled resources in the teleportation of single-mode coherent input states.
Given the input two-mode entangled state $\chi_{12}(x_1, p_1; x_2, p_2)$
and the two-mode entangled resource $\chi_{34}(x_3, p_3; x_4, p_4)$,
we compute the two-mode entangled output (swapped) state's characteristic function
$\chi_{out}^{(swapp)}(x_1, p_1;x_4, p_4)$, given by Eq.~(\ref{chiout}) for the realistic protocol
(or by Eq.~(\ref{chioutideal}) for the ideal protocol).
Such a two-mode entangled state is then used as a resource for the ideal teleportation protocol of single-mode coherent input states.
In summary:

- We first compute the output state of the swapping protocol
$\chi_{out}^{(swapp)}(x_1, p_1;x_4, p_4)$ associated with the entangled input state $X$
swapped with the entangled resource $Y$.

- Second, we compute the single-mode (teleported) state of the teleportation protocol
$\chi_{out}^{(telep)}(x_4, p_4)$:
\begin{eqnarray}
\chi_{out}^{(telep)}(x_{4},p_{4}) = \chi_{in}^{(coh)}(x_{4},p_{4})\, \chi_{out}^{(swapp)}(x_4,-p_4; x_4, p_4),\nonumber \\
&&
\label{MarianFormula}
\end{eqnarray}
where $\chi_{in}^{(coh)}(x_{4},p_{4})$ is the characteristic function of the coherent input state with displacement amplitude $\beta$.

- Third, we compute the fidelity of teleportation:
\begin{eqnarray}
\mathcal{F}_{X^{sw}Y} =\frac{1}{2\pi} \int dx_4 dp_4 \; \chi_{in}^{(coh)}(x_4, p_4) \chi_{out}^{(telep)}(-x_4, -p_4) \, , \nonumber \\
&&
\label{FidTelchi}
\end{eqnarray}
where the subscript $X^{sw}Y$ specifies the entangled states used as input $X$ and as resource $Y$
of the swapping protocol.

- Finally, we optimize the fidelity with respect to the available free parameters, and we use
the optimized fidelity to quantify the efficiency of the swapping protocol.

In Eq.~(\ref{FidTelchi}) the input state $X$ and the resource $Y$ can be any among the $TB$, $PS$, or $SB$ states.
With currently available technology, we have a more or less on-demand availability of Gaussian $TB$ states with finite squeezing,
while efficient production of non-Gaussian states with sizeable entanglement is more demanding.
Therefore, we may assume to have many copies of $TB$ states and few copies of $SB$ states.
With such a constraint, the most convenient approach would be to swap the non-Gaussian entanglement,
and thus to use $SB$ states as input states and $TB$ Gaussian states as resources.
For instance, in a long-distance communication scheme the entanglement swapping and entanglement purification protocols
can be performed to transfer non-Gaussian entanglement along a quantum channel divided into several segments.
If the above constraint could be removed in the next future, one would have on-demand availability also of $SB$ states.
In this desirable instance one could use $SB$ states both as input and as resources of the swapping protocol.
Therefore, for the sake of completeness, we compute the general form of the fidelity $\mathcal{F}_{SB^{sw}SB}$.
In this way, since $SB$ states include also $TB$ and $PS$ states for specific choices of the parameters,
we obtain as particular cases all the fidelities of interest,
i.e. $\mathcal{F}_{SB^{sw}TB}$, $\mathcal{F}_{PS^{sw}TB}$, and $\mathcal{F}_{TB^{sw}TB}$.

It is to verify that the optimal values for the phases $\phi_{hk}$ and $\theta_{hk}$ are
$\phi_{hk}=\pi$ and $\theta_{hk}=0$.
With such a choice, the dependence of the fidelity $\mathcal{F}_{SB^{sw}SB}$ on the two gains $g_1$ and $g_4$
reduces the dependence on the unique parameter $\tilde{g} = g_1 + g_4$, that can be exploited as the only gain parameter to be optimized.

The optimized fidelities are defined as:
\begin{equation}
\mathcal{F}_{X^{sw}Y}^{(opt)} = \max_{\mathcal{P}} \mathcal{F}_{X^{sw}Y} \,,
\end{equation}
where $\mathcal{P}$ denotes the set of free parameters available for optimization.
In the most general case in which generic $SB$ states are swapped with generic $SB$ resources,
the available free parameters for optimization are $\mathcal{P} = \{\delta_{12}, \delta_{34}, \tilde{g} \}$.

\section{Results}
\label{secResults}

In order to assess the teleportation performance when using non-Gaussian $SB$ resources, it is
convenient to introduce the relative fidelity, defined as~\cite{CVTelepNoi}:
\begin{equation}
\Delta \mathcal{F}_{SB}^{(X)} = \frac{\mathcal{F}_{SB}^{(opt)}-\mathcal{F}_{X}^{(ref)}}{\mathcal{F}_{X}^{(ref)}} \,,
\label{relfidelity}
\end{equation}
where $\mathcal{F}_{SB}^{(opt)}$ is the optimized fidelity
of teleportation associated with a non-Gaussian $SB$ resource
and $\mathcal{F}_{X}^{(ref)}$ is the (optimized) fidelity
associated to a reference resource $X$.

\begin{figure}[h]
  \centering
  \includegraphics[width=8cm]{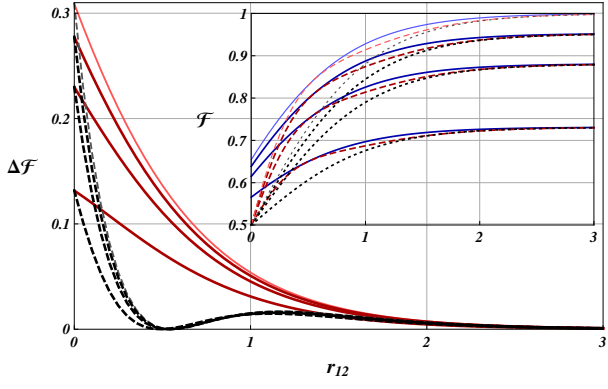}
  \caption{(Color online) Relative teleportation fidelities $\Delta \mathcal{F}_{SB^{sw}TB}^{(Y^{sw}TB)}$,
  as functions of the input squeezing parameter $r_{12}$ of the entangled input state. The squeezing parameter $r_{34}$
  of the swapping Gaussian $TB$ resource is fixed at three different values $r_{34} = 1.5, \, 1.0, \, 0.5$. The corresponding
  three curves are ordered from top to bottom at $r_{12}=0$. Full red curves: case $Y=TB$. Dashed black curves: case $Y=PS$.
  For comparison, we plot also the relative fidelities $\Delta \mathcal{F}_{SB}^{(X)}$
  associated with the corresponding non-swapped resources ($r_{34} \rightarrow \infty$),
  drawn in the same plot style, but with tinier line and lighter color.
  Inset: optimized absolute fidelities of teleportation $\mathcal{F}_{X^{sw}TB}^{(opt)}$,
  with $X=SB$ (full blue line), $X=PS$ (dashed red line), and $X=TB$ (dotted black line),
  as functions of $r_{12}$. For each resource, the curves corresponding, respectively, to
  $r_{34} = 1.5, \,  1, \, 0.5$ are ordered from top to bottom.
  For comparison, we include the optimized fidelities $\mathcal{F}_{X}^{(opt)}$
  associated with the corresponding non-swapped resources: $r_{34}\rightarrow \infty$.
  The corresponding curves are drawn in the same plot style, but with tinier line and lighter color.
  Fidelities associated with non-swapped resources saturate to unity.
  Fidelities associated with swapped resources saturate to a lower level,
  depending on the values of $r_{34}$.}
  \label{FigDue}
\end{figure}

Analogously, in order to quantify the teleportation performance
when using swapped non-Gaussian $SB$ resources with respect to reference swapped resources,
we generalize Eq.~(\ref{relfidelity}):
\begin{equation}
\Delta \mathcal{F}_{SB^{sw}X}^{(Y^{sw}Z)} =
\frac{\mathcal{F}_{SB^{sw}X}^{(opt)}-\mathcal{F}_{Y^{sw}Z}^{(ref)}}{\mathcal{F}_{Y^{sw}Z}^{(ref)}} \,,
\label{relfidelityswap}
\end{equation}
where $\mathcal{F}_{SB^{sw}X}^{(opt)}$ is the optimized fidelity of teleportation
associated with a $SB$ resource swapped with a resource $X$,
and $\mathcal{F}_{Y^{sw}Z}^{(ref)}$ is the reference (optimized) fidelity of teleportation
associated with a resource $Y$ swapped with a resource $Z$.

\subsection{Ideal swapping protocol}
\label{subsecIdealSwappPlots}

For the ideal swapping protocol one has $R_2 = R_3 = 0$, $\tau_1 = \tau_4 =0$.
First, we study the behavior of the teleportation fidelity
associated with different entangled resources ($X = SB, PS, TB$) swapped with Gaussian $TB$ resources ($Y = TB$).
In particular, we analyze the behavior of the relative fidelities $\Delta \mathcal{F}_{SB^{sw}TB}^{(TB^{sw}TB)}$
and $\Delta \mathcal{F}_{SB^{sw}TB}^{(PS^{sw}TB)}$.
We report them in Fig.~\ref{FigDue} as functions
of the squeezing parameter $r_{12}$ of the swapped resource, for different fixed values of the swapping squeezing parameter $r_{34}$.

The relative improvement in the fidelity of teleportation that is obtained using swapped $SB$ resources
increases for growing $r_{34}$ and equals that of non-swapped resources
at sufficiently large values of $r_{34}$. A particularly significant improvement is obtained over the Gaussian $TB$ instance.
Furthermore, the use of swapped $SB$ resources improves the teleportation fidelity also when compared to the use of swapped $PS$ resources,
especially for values of the two-mode squeezing $r_{12} \in [0,1]$.

\begin{figure}[h]
  \centering
  \includegraphics[width=8cm]{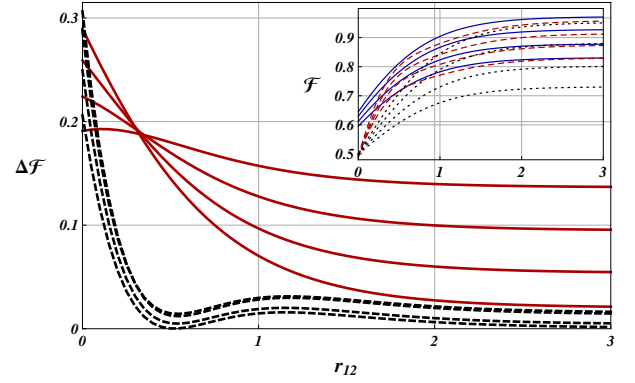}
  \caption{(Color online) Relative teleportation fidelities $\Delta \mathcal{F}_{SB^{sw}SB}^{(X^{sw}X)}$
  as functions of the input squeezing parameter $r_{12}$ of the entangled input state. The squeezing parameter $r_{34}$
  of the swapping resource is fixed at four different values $r_{34} = 1.5, \, 1.0, \, 0.7, \, 0.5$. The corresponding
  four curves are ordered from top to bottom at $r_{12}=0$. Full red curves: case $X=TB$. Dashed black curves: case $X=PS$.
  For $X=TB$, the use of non-Gaussian $SB$ swapping resources guarantees a finite non-vanishing improvement in the teleportation
  fidelity for any value of the input squeezing $r_{12}$.
  Inset: optimized absolute fidelities of teleportation $\mathcal{F}_{X^{sw}TB}^{(opt)}$,
  with $X=SB$ (full blue line), $X=PS$ (dashed red line), and $X=TB$ (dotted black line),
  as functions of $r_{12}$. For each resource, the curves corresponding, respectively, to
  $r_{34} = 1.5, \,  1, \, 0.7, \, 0.5$ are ordered from top to bottom.}
  \label{FigSei}
\end{figure}

In the inset of Fig.~\ref{FigDue}, we report the optimized fidelities
$\mathcal{F}_{SB^{sw}TB}^{(opt)}$, $\mathcal{F}_{PS^{sw}TB}^{(opt)}$
and $\mathcal{F}_{TB^{sw}TB}^{(opt)}$
as functions of $r_{12}$, at the same different fixed values of $r_{34}$.
For comparison, in the same inset we report also the corresponding optimized fidelities
$\mathcal{F}_{SB}^{(opt)}$, $\mathcal{F}_{PS}^{(opt)}$, and $\mathcal{F}_{TB}^{(opt)}$
associated with the same non-swapped resources (equivalently $r_{34} \rightarrow\infty$).
For a fixed finite value of $r_{34}$ the fidelities $\mathcal{F}_{X^{sw}TB}^{(opt)}$
are always lower than the ideal ones $\mathcal{F}_{X}^{(opt)}$;
as expected, for large values of the swapping squeezing strength
$r_{34}$, the fidelities $\mathcal{F}_{X^{sw}TB}^{(opt)}$ tend to the ideal fidelities.
The saturation level that is featured at large values of the squeezing of the swapped resource $r_{12}$,
is higher and tends to the ideal value one for larger fixed values of $r_{34}$.

We now consider the case in which both the input state and the swapping resource are non-Gaussian, and
we thus study the optimized fidelities $\mathcal{F}_{X^{sw}X}^{(opt)}$ with $X=SB,\,PS,\,TB$.
In this instance, although it is possible to obtain the exact analytical expressions for each $\mathcal{F}_{X^{sw}X}$,
their optimization over the set of free parameters must be carried out numerically.

In Fig.~\ref{FigSei} we report the relative teleportation fidelities
$\Delta \mathcal{F}_{SB^{sw}SB}^{(TB^{sw}TB)}$ and $\Delta \mathcal{F}_{SB^{sw}SB}^{(PS^{sw}PS)}$
as functions of the squeezing parameter $r_{12}$ of the input entangled state, for different values
of the squeezing parameter $r_{34}$ of the swapping resource. For the relative fidelity $\Delta \mathcal{F}_{SB^{sw}SB}^{(TB^{sw}TB)}$,
the curves corresponding to different values of $r_{34}$ intersect at $r_{12} \simeq 0.4$.

The relative fidelities feature maximal enhancement when using swapped non-Gaussian $SB$ resources, especially
with respect to the use of swapped Gaussian $TB$ resources. Remarkably, the use of non-Gaussian $SB$ swapping resources guarantees a constant,
finite and non-vanishing improvement in the teleportation fidelity for sufficiently large values of the input squeezing $r_{12}$.
Indeed, in the complete non-Gaussian instance the optimized (swapped) $SB$ resources never collapse
onto the optimized (swapped) $PS$ resources. Correspondingly, the relative fidelity {\em never vanishes}.

\begin{figure}[h]
  \centering
  \includegraphics[width=8cm]{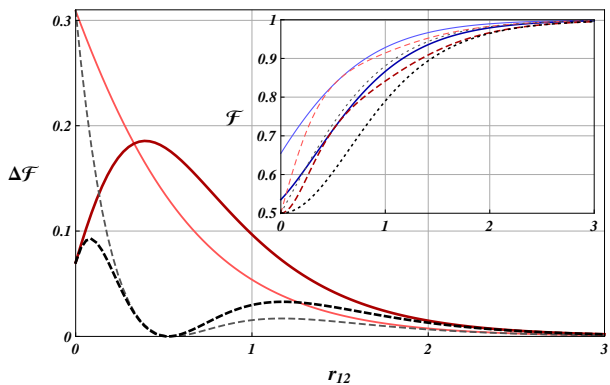}
  \caption{(Color online) Relative teleportation fidelities $\Delta \mathcal{F}_{SB^{sw}SB}^{(X^{sw}X)}$
  as functions of the input squeezing parameter $r_{12}$, in the fully symmetric instance
  $r_{34} = r_{12}$, $\delta_{34} = \delta_{12}$. Full red curve: case $X = TB$. Dashed black curve:
  case $X = PS$. For comparison, we also report the relative fidelities $\Delta \mathcal{F}_{SB}^{(X)}$
  associated with the corresponding non-swapped resources $(r_{34}\rightarrow \infty)$,
  drawn in the same plot style, but with tinier line and lighter color.
  Inset: optimized absolute fidelities of teleportation $\mathcal{F}_{X^{sw}TB}^{(opt)}$
  as functions of $r_{12}$, with $X=SB$ (full blue line), $X=PS$ (dashed red line), and $X=TB$ (dotted black line).
  For comparison, we also report the ideal fidelities $\mathcal{F}_{X}^{(opt)}$
  associated with the corresponding non-swapped resources ($r_{34}\rightarrow \infty$), drawn with
  the same plot style, but with tinier line and lighter color.}
  \label{FigSeibis}
\end{figure}

Finally, we consider a situation in which a certain number of identical copies of entangled resource states is available.
Such case allows for a minimization of the experimental costs
required for the generation of the same resources and for the optimization of the experimentally tunable free parameters.
In this instance, the swapping resources are identical to the input states of the swapping protocol, so that $r_{34}=r_{12}$, $\delta_{34}=\delta_{12}$ for the most general case of $SB$ resources.
In Fig.~\ref{FigSeibis}, we report the relative teleportation fidelities $\Delta \mathcal{F}_{SB^{sw}SB}^{(X^{sw}X)}$
as functions of the squeezing parameter $r_{12}=r_{34}$, optimized over the free parameters $\delta_{12}=\delta_{34}$.

Summing up, the ideal case non-Gaussian $SB$ resources always outperform both Gaussian $TB$ and non-Gaussian $PS$ resources at small values of the input two-mode squeezing $r_{12}$. The relative teleportation fidelities decrease and vanish asymptotically with arbitrarily increasing values of the two-mode squeezing $r_{12}$, when using Gaussian $TB$ swapping resources (see Fig.~\ref{FigDue}). Indeed, in the infinite squeezing limit, both Gaussian $TB$ and non-Gaussian $SB$ resources approach the ideal $EPR$ state, yielding unit teleportation fidelity. The same asymptotic behavior occurs
in the symmetric case, for which the swapping resources coincide with the input entangled states (see Fig.~\ref{FigSeibis}).

In the case of non-Gaussian resources on demand, the advantage in using $SB$ states persists for any value of the input
two-mode squeezing $r_{12}$, asymptotically yielding a constant, finite and non-vanishing improvement in the performance of the
entanglement swapping protocol (see Fig.~\ref{FigSei}).

\subsection{Realistic swapping protocol}

Here we investigate the relative performance of Gaussian and non-Gaussian swapped resources in a realistic swapping protocol.
We consider the situation of complete prior knowledge of the experimental parameters describing losses, imperfections, and decoherence effects.
From an operational point of view this corresponds to the complete characterization of
the experimental apparatus, including the inefficiencies of the photo-detectors,
the lengths, and the damping rates of the noisy channels.

\begin{figure}[h]
  \centering
  \includegraphics[width=8cm]{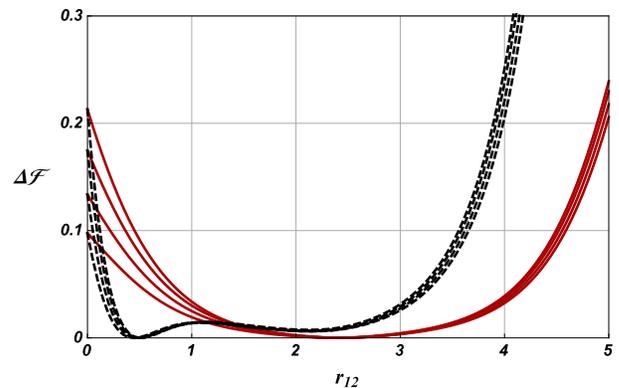}
  \caption{(Color online)
  Relative teleportation fidelities $\Delta \mathcal{F}_{SB^{sw}TB}^{(Y^{sw}TB)}$
  with $Y=TB$ (full line) and $Y=PS$ (dashed line),
  as functions of the squeezing parameter $r_{12}$ of the swapped input state.
  The squeezing parameter of the swapping $TB$ resource is fixed at the values $r_{34} =0.5, \; 0.7, \; 1, \; 1.5$,
  corresponding to the curves ordered at $r_{12}=0$ from bottom to top.
  The parameters of the experimental apparatus are fixed as:
  $\tau_1 =0.1$, $n_{th,1}=0$, $\tau_4 =0.2$, $n_{th,4}=0$, $R_2 =\sqrt{0.05}$, $R_3 =\sqrt{0.05}$.
  In the inset are plotted optimized fidelities of teleportation $\mathcal{F}_{X^{sw}TB}^{(opt)}$,
  with $X=SB$ (full line), $X=PS$ (dashed line), and $X=TB$ (dotted line),
  as functions of $r_{12}$. The curves corresponding to a given resource (same plot style)
  are ordered from both to top for growing $r_{34}=0.5, \; 0.7, \; 1, \; 1.5$.}
  \label{FigSette}
\end{figure}

In the following, we proceed as in the previous subsection. In particular, we carry out the optimization of the fidelities
once the values of the parameters associated with the experimental apparatus have been fixed.

In Fig.~\ref{FigSette} we report the relative teleportation fidelities $\Delta \mathcal{F}_{SB^{sw}TB}^{(Y^{sw}TB)}$
for $Y \, = \, TB \, , \, PS$, as functions of the input two-mode squeezing $r_{12}$. In the inset we also report the
optimized absolute fidelities of teleportation.

For sufficiently small values of the two-mode squeezing $r_{12}$ the behavior of the relative teleportation fidelities is analogous
to that of the same quantities in the ideal instance (see Fig.~\ref{FigDue}).

The behavior changes dramatically at intermediate and large values of $r_{12}$.
For growing input squeezing $r_{12}$, decoherence affects more and more severely the quality of the swapping resources, as an
initially larger number of squeezed photons feeding the lossy channel is rapidly converted into a larger number of incoherent,
thermal photons.

\begin{figure}[h]
  \centering
  \includegraphics[width=8cm]{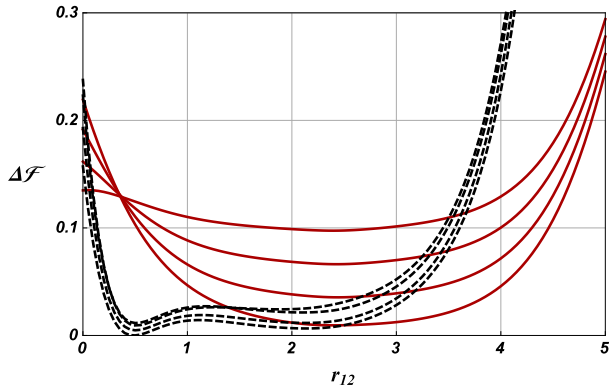}
  \caption{(Color online) Relative teleportation fidelities $\Delta \mathcal{F}_{SB^{sw}SB}^{(X^{sw}X)}$
  with $X=TB$ (full line) and $X=PS$ (dashed line),
  as functions of the squeezing parameter $r_{12}$ of the swapped input state.
  The squeezing parameter of the swapping resource is fixed at the values $r_{34} =0.5, \; 0.7, \; 1, \; 1.5$,
  corresponding to the curves ordered at $r_{12}=0$ from bottom to top.
  The parameters of the experimental apparatus are fixed as in Fig.~\ref{FigSette}.
  In the instance $X=TB$, the curves associated with different $r_{34}$ show an intersection at a certain $r_{12}$.
  In the inset are plotted optimized fidelity of teleportation $\mathcal{F}_{X^{sw}TB}^{(opt)}$,
  with $X=SB$ (full line), $X=PS$ (dashed line), and $X=TB$ (dotted line),
  as functions of $r_{12}$. The curves corresponding to a given resource (same plot style)
  are ordered from bottom to top for growing $r_{34}=0.5, \; 0.7, \; 1, \; 1.5$.}
  \label{FigOtto}
\end{figure}

However, decoherence affects differently the different resources. The strongest deterioration is felt by the Gaussian
$TB$ resource and by the non-Gaussian $PS$ resource, while the non-Gaussian $SB$ resource turns out to be more resilient. Indeed, as shown in Fig.~\ref{FigSette}, the use of $SB$ entangled states allows for a relative teleportation fidelity that is monotonically improving with increasing two-mode squeezing $r_{12}$.

\begin{figure}[h]
  \centering
  \includegraphics[width=8cm]{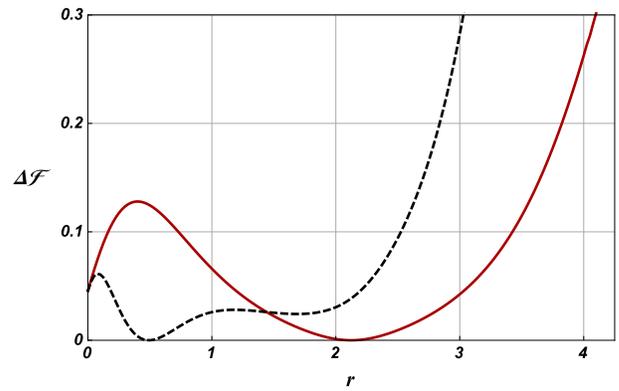}
  \caption{(Color online) Relative teleportation fidelities $\Delta \mathcal{F}_{SB^{sw}SB}^{(X^{sw}X)}$
  with $X=TB$ (full line) and $X=PS$ (dashed line), as functions of the squeezing parameter $r_{12}$,
  in the fully symmetric instance, i.e. $r_{34}=r_{12}$, $\delta_{34}=\delta_{12}$.
  In the inset are plotted optimized fidelities of teleportation $\mathcal{F}_{X^{sw}TB}^{(opt)}$,
  with $X=SB$ (full line), $X=PS$ (dashed line), and $X=TB$ (dotted line),
  as functions of $r_{12}$.
  The parameters of the experimental apparatus are fixed as in Fig.~\ref{FigSette}.}
  \label{FigNove}
\end{figure}

Resilience against decoherence effects with the use of $SB$ resources is even more striking under the assumption of non-Gaussian entangled states available on demand, as shown in Fig.~\ref{FigOtto}. A similar behavior is observed when one considers also the symmetric case, when the swapping resources are equal to the input ones, as reported in Fig.~\ref{FigNove}.

The highest achievable values of two-mode squeezing in the optical regime are currently limited to $r_{12} \simeq 2$ \cite{Schnabel}. In the near future, foreseeable advance towards the experimental accessibility of higher squeezing values and capability of routinely generating non-Gaussian resources by linear optics and precise conditional measurements \cite{SqueezBellEngineer}, will make the use non-Gaussian $SB$ entangled resources quite compelling in order to control and reduce the disruptive effects of environmental decoherence.

\section{Conclusions}

The present paper is part of a wide investigation on the effectiveness of non-Gaussian resources for the
implementation of Quantum Information and Communication protocols.
This investigation has included till now
the introduction of a general class of non-Gaussian entangled resources
(the Squeezed Bell states) which encompasses many Gaussian and non-Gaussian entangled states \cite{CVTelepNoi},
the study of their efficiency in implementing Quantum Teleportation protocols in ideal and in realistic conditions \cite{CVTelepNoi,RealCVTelepNoi,CVSqueezTelepNoi},
and the proposal of a generating scheme for this class of states \cite{SqueezBellEngineer}.

By using the Squeezed Bell states as non-Gaussian entangled resources,
we investigated the efficiency of the vLB CV quantum swapping protocol
for the transmission of quantum  states and entanglement.
In order to evaluate the performance of the swapping protocol
we have exploited a criterion based on the ideal teleportation
of input coherent states using as entangled resources
the swapped states. In particular, the teleportation fidelity has been assumed
as a convenient indicator to quantify the performance levels.

Non-Gaussian Squeezed Bell resources
allow for optimization procedures, providing high values of the fidelities
both in the ideal and in the realistic instances. In realistic conditions, the tunable parameters measuring the degree of non-Gaussianity
of the squeezed Bell resources allow for an effective control of the decoherence effects
caused by losses and inefficiencies.

In all cases, we carried out a detailed comparison of the performance of optimized squeezed Bell resources with respect to the most significant currently available reference classes of entangled resources: Gaussian twin beams and non-Gaussian photon-subtracted squeezed states.
The bottom line of our study is that the use of Squeezed Bell entangled resources becomes compelling
in realistic conditions when going beyond the low-squeezing regime.

In future work, we will investigate the possibility of obtaining further improvements in the efficiency of teleportation and swapping protocols using non-Gaussian resources, by considering schemes that involve communication channels with diversified characteristics and varying sets of tunable parameters \cite{TelepZubairy}.

\section{Acknowledgments}

We acknowledge the EU FP7 Cooperation STREP Project EQuaM - Emulators of Quantum Frustrated Magnetism, Grant Agreement No. 323714. We also acknowledge financial support from the Italian Minister of Scientific Research (MIUR) under the national PRIN programme.

\appendix

\section{CV entanglement swapping protocol in the characteristic function representation}
\label{appsecCharFuncSwapp}

The characteristic function representation provides a most elegant and compact description of the CV teleportation protocol \cite{MarianCVTelep}.
Such description proves to be particularly convenient in the instance of non-Gaussian resources
\cite{CVTelepNoi,RealCVTelepNoi,CVTelepNoisyNoi,CVSqueezTelepNoi},
and has been generalized to include the non-ideal case \cite{RealCVTelepNoi}.
In this appendix, we apply this formalism to the description of the realistic CV entanglement swapping protocol,
schematically illustrated in Fig.~\ref{FigSwapping}.

Let $\rho_{12}^{A}$ and $\rho_{34}^{B}$ denote the density matrices associated with
the two-mode entangled input pure state of modes $1$ and $2$,
and the two-mode entangled pure resource of modes $3$ and $4$, respectively.
The global four-mode initial state is the biseparable state $\rho_{0}= \rho_{12}^{A} \otimes \rho_{34}^{B}$
and the corresponding initial global characteristic function $\chi_{0}$ associated with $\rho_{0}$ reads:
\begin{eqnarray}
\chi_{0}(\alpha_1;\alpha_2;\alpha_3;\alpha_4)=&& Tr[\prod_{j=1}^{4} D_j (\alpha_j) \rho_{0}] \nonumber \\
=&&
\chi_{12}(\alpha_1;\alpha_2) \; \chi_{34}(\alpha_3;\alpha_4) \,,
\label{chi0alpha}
\end{eqnarray}
where $Tr$ denotes the trace operation,
$D_j (\alpha_j)$ denotes the displacement operator of mode $j$ $(j=1,\ldots,4)$,
$\chi_{12}$ is the characteristic function of the two-mode input state,
and $\chi_{34}$ is the characteristic function of the two-mode resource.
By introducing the quadrature operators $X_{j}=\frac{1}{\sqrt{2}}(a_{j}+a_{j}^{\dag})$ and
$P_{j}=\frac{i}{\sqrt{2}}(a_{j}^{\dag}-a_{j})$,
and the corresponding phase space variables
$x_{j}= \frac{1}{\sqrt{2}}(\alpha _{j}+\alpha _{j}^{*})$ and
$p_{j}=\frac{i}{\sqrt{2}}( \alpha_{j}^{*}-\alpha _{j})$,
the characteristic function can be written in terms of $x_j$, $p_j$,
i.e. $\chi_{0}(\alpha_1;\alpha_2;\alpha_3;\alpha_4)\equiv \chi_{0}(x_1,p_1;x_2,p_2;x_3,p_3;x_4,p_4)$.

The first step of the protocol consists of a Bell measurement at the first user's location.
The modes $2$ and $3$ are mixed at a balanced beam splitter;
the effects of photon losses and the inefficiencies of the photodetectors are simulated
by two additional fictitious beam splitters placed in front of the detectors, characterized by the transmissivities
$T_{j}^2$ (reflectivity $R_{j}^2=1-T_{j}^2$), $j=2,3$.
Let us denote by $\tilde{x}$ and $\tilde{p}$ the homodyne measurements
of the first quadrature of the mode $3$ and of the second quadrature of the mode $2$, respectively.
The description of realistic Bell measurements using the formalism of the characteristic function is discussed in full detail
in Ref.~\cite{RealCVTelepNoi}.
Here we just provide the final expression of the characteristic function $\chi_{Bm}(x_1,p_1;x_4,p_4)$
associated with the entire measurement process:
\begin{eqnarray}
&&\chi_{Bm}(x_1,p_1;x_4,p_4) = \frac{\mathcal{P}^{-1}(\tilde{p},\tilde{x}) }{(2\pi)^{2}}
\int
d\xi d\upsilon \, e^{i\xi \tilde{p} - i \tilde{x} \upsilon }  \nonumber \\
&&\times \chi_{12} \left(x_1,p_1; \frac{T_2 \xi}{\sqrt{2}},\frac{T_3 \upsilon}{\sqrt{2}} \right)
\chi_{34} \left( \frac{T_2 \xi}{\sqrt{2}},-\frac{T_3 \upsilon}{\sqrt{2}}; x_4,p_4\right) \nonumber \\
&&\times \exp\left[-\frac{R_2^2}{4} \xi^2 -\frac{R_3^2}{4} \upsilon^2 \right] \,,
\label{chiBellmeas}
\end{eqnarray}
where the function $\mathcal{P}(\tilde{p},\tilde{x})$ is the distribution of the measurement outcomes
$\tilde{p}$ and $\tilde{x}$, that is:
\begin{eqnarray}
&&\mathcal{P}(\tilde{p},\tilde{x}) = \frac{1 }{(2\pi)^{2}}
\int
d\xi d\upsilon \, e^{i\xi \tilde{p} - i \tilde{x} \upsilon } e^{-\frac{R_2^2}{4} \xi^2 -\frac{R_3^2}{4} \upsilon^2} \nonumber \\
&&\times \chi_{12} \left(0,0; \frac{T_2 \xi}{\sqrt{2}},\frac{T_3 \upsilon}{\sqrt{2}} \right)
\chi_{34} \left( \frac{T_2 \xi}{\sqrt{2}},-\frac{T_3 \upsilon}{\sqrt{2}};0,0\right) .
\end{eqnarray}

After measurement, modes $1$ and $4$ propagate in noisy channels (e.g. optical fibers)
towards Alice's and Bob's locations, respectively.
The dynamics of a multimode system subject to decoherence is described, in the interaction picture,
by the following master equation for the density operator $\rho$ \cite{WallsMilburn,DecohReview}:
\begin{equation}
\partial_{t} \rho \,=\, \sum_{i=1,4} \frac{\Upsilon_i}{2}
\left\{ n_{th,i} L[a_{i}^{\dag}] \rho + (n_{th,i}+1) L[a_{i}] \rho \right\} \,,
\label{MasterEq}
\end{equation}
where the Lindblad superoperators are defined as $L[\mathcal{O}]
\rho \equiv 2 \mathcal{O} \rho \mathcal{O^{\dag}}
- \mathcal{O^{\dag}} \mathcal{O} \rho - \rho \mathcal{O^{\dag}} \mathcal{O}$,
$\Upsilon_i$ is the mode damping rate, and $n_{th,i}$ is the number of thermal photons in mode $i$.
Because of decoherence due to propagation in the noisy channels,
the characteristic function (\ref{chiBellmeas}) can be rewritten in the following form:
\begin{eqnarray}
&&\chi_{t}(x_{1},p_{1};x_{4},p_{4}) \,=\, \nonumber \\
&&
\chi_{Bm}(e^{-\frac{1}{2}\Upsilon_1 t}x_{1},e^{-\frac{1}{2}\Upsilon_1 t}p_{1};e^{-\frac{1}{2}\Upsilon_4 t}x_{4},e^{-\frac{1}{2}\Upsilon_4 t}p_{4}) \nonumber \\
&&\times \; e^{-\frac{1}{2} \sum_{i=1,4} (1-e^{-\Upsilon_i t})\left(\frac{1}{2}+n_{th,i}\right)(x_{i}^{2}+p_{i}^{2})}. \label{solchit}
\end{eqnarray}

The description of the technical features of the experimental apparatus,
e.g. the efficiency of the photodetectors, and characteristics as the length of the channels (fibers), and
the temperature of the environment, is complete once the quantities
$T_{j}$ (equivalently $R_{j}$, $j=2,3$), $\Upsilon_i$, and $n_{th,i}$ ($i=1,4$) are specified and
fixed at certain given values.

In the last step of the protocol, two local unitary displacements $\lambda_1$ and $\lambda_4$ are performed at Alice's and Bob's locations;
a local unitary displacement $\lambda_1 = - g_1 ( \tilde{x} - i\tilde{p})$ is performed on mode $1$,
and a local unitary displacement $\lambda_4 = g_4 ( \tilde{x} + i\tilde{p})$ is performed on mode $4$.
The real parameters $g_1$ and $g_4$ denote the gain factors of the displacement transformations \cite{TelepGainBowen}.
After such local unitary operations, the characteristic function reads:
\begin{eqnarray}
\chi_D(x_1,p_1;x_4,p_4) = && e^{-i\sqrt{2}\tilde{x}(g_1 p_1 - g_4 p_4)-i\sqrt{2}\tilde{p}(g_1 x_1 + g_4 x_4)}
 \nonumber \\
&& \times \chi_{t}(x_1,p_1;x_4,p_4)  \,.
\label{chiDisplaced}
\end{eqnarray}

Finally, in order to obtain the output characteristic function $\chi_{out}(x_1,p_1;x_4,p_4)$,
describing the output two-mode entangled state of the entanglement swapping protocol,
one must take the average over all the possible outcomes $\tilde{p}$ and $\tilde{x}$ of the Bell measurements:
\begin{equation}
\chi_{out}^{(swapp)}(x_1,p_1;x_4,p_4) = \int d\tilde{x} d\tilde{p} \mathcal{P}(\tilde{p},\tilde{x})\chi_D(x_1,p_1;x_4,p_4),
\label{chiout2}
\end{equation}
where $\tau_i = \Upsilon_i t$.
The above integral yields the final expression (\ref{chiout}) for the characteristic function
associated with the swapped resource.

The core mathematical task is thus the explicit evaluation
of the characteristic function $\chi_{out}^{(swapp)}(x_1,p_1;x_4,p_4)$
associated with the output of the realistic swapping protocol, as expressed by Eq.~(\ref{chiout}).
We have determined its analytical expression for the most general non-Gaussian setting,
that is the entanglement swapping of $SB$ input states using $SB$ states as resources.
As the class of $SB$ states contains as special cases both the Gaussian $TB$ states and the
non-Gaussian $PS$ states, the general expression of the output characteristic function reduces to
the explicit expression for these special Gaussian and non-Gaussian cases as well.
We do not report here the explicit analytical expression of $\chi_{out}^{(swapp)}(x_1,p_1;x_4,p_4)$,
as it is exceedingly long and cumbersome and does not yield any particularly useful physical insight.
On the other hand, having obtained the explicit expression of the output characteristic function of the swapping protocol, it is straightforward to compute
the output characteristic function $\chi_{out}^{(telep)}(x_{4},p_{4})$ of the subsequent ideal teleportation protocol,
Eq.~(\ref{MarianFormula}).

Finally, we have derived the analytical expression for the teleportation fidelity $\mathcal{F}_{X^{sw}Y}$,
Eq.~(\ref{FidTelchi}), which in the most general instance is $\mathcal{F}_{SB^{sw}SB}$.
Such a fidelity depends on the following parameters:
the squeezing amplitudes and phases $r_{12}$, $\phi_{12}$, $r_{34}$, $\phi_{34}$
and the angles and phases $\delta_{12}$, $\theta_{12}$, $\delta_{34}$, $\theta_{34}$
of the input states and of the resources; the parameters associated with the experimental apparatus
are listed in Tab.~\ref{tableExpPara}.
Without loss of generality, as already verified in Refs.\cite{CVTelepNoi,RealCVTelepNoi}, one can obtain some
significant simplifications by fixing the phases of the $SB$ states.
Specifically, we set the non-Gaussian phases $\theta_{12}=\theta_{34}=0$ and the squeezing phases $\phi_{12}=\phi_{34}=\pi$
in Eq.~(\ref{squeezBell}). With such choice, the teleportation fidelity depends
on the two gains $g_i$ $(i=1,4)$ through the total gain parameter $\tilde{g}=g_1+g_4$,
both in the ideal and in the realistic protocols.


\begin{thebibliography}{99}

\bibitem{QCommunGisin}
N. Gisin, G. Ribordy, W. Tittel, and H. Zbinden, Rev. Mod. Phys. {\bf 74}, 145 (2002).

\bibitem{Qrepet}
H. J. Briegel, W. Dur, J. I. Cirac, and P. Zoller, Phys. Rev. Lett. {\bf 81}, 5932 (1998).

\bibitem{QICV}
S. L. Braunstein and P. van Loock, Rev. Mod. Phys. {\bf 77}, 513 (2005).

\bibitem{QKDSanders}
A. Khalique and B. C. Sanders, J. Opt. Soc. Am. B {\bf 32}, 2382 (2015).

\bibitem{QKDZippilli}
M. Asjad, S. Zippilli, P. Tombesi, and D. Vitali, Phys. Scr. {\bf 90}, 074055 (2015).

\bibitem{vanLoockBraunstein}
P. van Loock and S. L. Braunstein, Phys. Rev. A {\bf 61}, 010302 (1999).

\bibitem{ExpSwap1}
X. Jia, X. Su, Q. Pan, J. Gao, C. Xie, and K. Peng, Phys. Rev. Lett. {\bf 93}, 250503 (2004).

\bibitem{ExpSwap2}
T. Yang, Q. Zhang, T.-Y. Chen, S. Lu, J. Yin, J.-W. Pan, Z.-Y. Wei, J.-R. Tian, and J. Zhang,
Phys. Rev. Lett. {\bf 96}, 110501 (2006).

\bibitem{ExpSwap3}
R.-B. Jin, M. Takeoka, U. Takagi, R. Shimizu, and M. Sasaki, Scientific Reports {\bf 5}, 9333 (2015).

 \bibitem{HybridSwap}
S. Takeda, M. Fuwa, P. van Loock, and A. Furusawa, Phys. Rev. Lett. {\bf 114}, 100501 (2015).

\bibitem{OptimalGaussSwapp}
J. Hoelscher-Obermaier and P. van Loock, Phys. Rev. A {\bf 83}, 012319 (2011).

\bibitem{CVTelepNoi}
F. Dell'Anno, S. De Siena, L. Albano, and F. Illuminati, Phys. Rev. A {\bf 76}, 022301 (2007).

\bibitem{RealCVTelepNoi}
F. Dell'Anno, S. De Siena, and F. Illuminati, Phys. Rev. A {\bf 81}, 012333 (2010).

\bibitem{CVTelepNoisyNoi}
F. Dell'Anno, S. De Siena, L. Albano, and F. Illuminati, Eur. Phys. J. Special Topics {\bf 160}, 115 (2008).

\bibitem{CVSqueezTelepNoi}
F. Dell'Anno, S. De Siena, G. Adesso, and F. Illuminati, Phys. Rev. A {\bf 82}, 062329 (2010).

\bibitem{KimBS}
M. S. Kim, W. Son, V. Bu\v{z}ek, and P. L. Knight, Phys. Rev. A {\bf 65}, 032323 (2002).

\bibitem{DodonovDisplnumb}
V. V. Dodonov and L. A. de Souza, J. Opt. B: Quantum Semiclass. Opt. {\bf 7}, S490 (2005).

\bibitem{Cerf}
N. J. Cerf, O. Kr\"uger, P. Navez, R. F. Werner, and M. M. Wolf, Phys. Rev. Lett. {\bf 95}, 070501 (2005).

\bibitem{Opatrny}
T. Opatrn\'y, G. Kurizki, and D.-G. Welsch, Phys. Rev. A {\bf 61}, 032302 (2000).

\bibitem{Cochrane}
P. T. Cochrane, T. C. Ralph, and G. J. Milburn, Phys. Rev. A {\bf 65}, 062306 (2002).

\bibitem{Olivares}
S. Olivares, M. G. A. Paris, and R. Bonifacio, Phys. Rev. A {\bf 67}, 032314 (2003).

\bibitem{KitagawaPhotsub}
A. Kitagawa, M. Takeoka, M. Sasaki, and A. Chefles, Phys. Rev. A {\bf 73}, 042310 (2006).

\bibitem{YangLi}
Y. Yang and F.-L. Li, Phys. Rev. A {\bf 80}, 022315 (2009).

\bibitem{QEstimNoi}
G. Adesso, F. Dell'Anno, S. De Siena, F. Illuminati, and L. A. M. Souza, Phys. Rev. A {\bf 79}, 040305(R) (2009).

\bibitem{QCMenicucci}
N. C. Menicucci, P. van Loock , M. Gu, C. Weedbrook, T. C. Ralph, and M. A. Nielsen, Phys. Rev. Lett. {\bf 97}, 110501 (2006).

\bibitem{NGBartley}
T. J. Bartley and I. A. Walmsley, New J. Phys. {\bf 17}, 023038 (2015).

\bibitem{NonGaussTelepChina}
S. Wang, L.-L. Hou, X.-F. Chen, and X.-F. Xu, Phys. Rev. A {\bf 91}, 063832 (2015).

\bibitem{ExtremalGaussian}
M. M. Wolf, G. Giedke, and J. I. Cirac, Phys. Rev. Lett. {\bf 96}, 080502 (2006).

\bibitem{Genoni}
M. G. Genoni and M. G. A. Paris, Phys. Rev. A {\bf 82}, 052341 (2010).

\bibitem{Scheel}
J. Eisert, S. Scheel, and M. B. Plenio, Phys. Rev. Lett. {\bf 89}, 137903 (2002).

\bibitem{EisertLimitations}
M. Ohliger, K. Kieling, and J. Eisert, Phys. Rev. A {\bf 82}, 042336 (2010).

\bibitem{PhysRep}
F. Dell'Anno, S. De Siena, and F. Illuminati, Phys. Rep. {\bf 428}, 53 (2006).

\bibitem{CxKerrKorolkova}
T. Tyc and N. Korolkova, New J. Phys. {\bf 10}, 023041 (2008).

\bibitem{NonlinBogoNoi}
F. Dell'Anno, S. De Siena, and F. Illuminati, Phys. Rev. A {\bf 69}, 033812 (2004);
{\it ibidem} {\bf 69}, 033813 (2004).

\bibitem{AgarTara}
G. S. Agarwal and K. Tara, Phys. Rev. A {\bf 43}, 492 (1991).

\bibitem{DeGauss1}
G. Bjork and Y. Yamamoto, Phys. Rev. A {\bf 37}, 4229 (1988).

\bibitem{DeGauss2}
Z. Zhang and H. Fan, Phys. Lett. A {\bf 165}, 14 (1992).

\bibitem{DeGauss3}
M. Dakna, T. Anhut, T. Opatrn\'{y}, L. Kn\"{o}ll, and D. G. Welsch, Phys. Rev. A {\bf 55}, 3184 (1997).

\bibitem{DeGauss4}
M. S. Kim, E. Park, P. L. Knight, and H. Jeong, Phys. Rev. A {\bf 71}, 043805 (2005).

\bibitem{DeGauss5}
D. Menzies and R. Filip, Phys. Rev. A {\bf 79}, 012313 (2009).

\bibitem{DeGauss6}
S.-Y. Lee and H. Nha, Phys. Rev. A {\bf 82}, 053812 (2010).

\bibitem{DeGauss7}
M. G. Genoni, F. A. Beduini, A. Allevi, M. Bondani, S. Olivares, and M. G. A. Paris,
Phys. Scr. {\bf T140}, 014007 (2010).

\bibitem{DeGauss8}
S.-Y. Lee, S.-W. Ji, H.-J. Kim, and H. Nha, Phys. Rev. A {\bf 84}, 012302 (2011).

\bibitem{DeGauss9}
X.-X. Xu, H.-C. Yuan, and H.-Y. Fan, J. Opt. Soc. Am. B {\bf 32}, 1146 (2015).

\bibitem{ZavattaScience}
A. Zavatta, S. Viciani, and M. Bellini, Science {\bf 306}, 660 (2004).

\bibitem{ExpdeGauss1}
A. I. Lvovsky and S. A. Babichev, Phys. Rev. A {\bf 66}, 011801 (2002).

\bibitem{ExpdeGauss2}
J. Wenger, R. Tualle-Brouri, and P. Grangier, Phys. Rev. Lett. {\bf 92}, 153601 (2004).

\bibitem{Solimeno1}
V. D'Auria, A. Chiummo, M. De Laurentis, A. Porzio, S. Solimeno, and M. G. A. Paris,
Opt. Expr. {\bf 13} 948 (2005).

\bibitem{Grangier}
A. Ourjoumtsev, A. Dantan, R. Tualle-Brouri, and P. Grangier, Phys. Rev. Lett. {\bf 98}, 030502 (2007).

\bibitem{BelliniProbing}
V. Parigi, A. Zavatta, M. Kim, and M. Bellini, Science {\bf 317}, 1890 (2007).

\bibitem{GrangierCats}
A. Ourjoumtsev, H. Jeong, R. Tualle-Brouri, and P. Grangier, Nature {\bf 448}, 784 (2007).

\bibitem{Solimeno2}
V. D'Auria, C. de Lisio, A. Porzio, S. Solimeno, J. Anwar, and M. G. A. Paris,
Phys. Rev. A {\bf 81}, 033846 (2010).

\bibitem{ExpEtesse}
J. Etesse, M. Bouillard, B. Kanseri, and R. Tualle-Brouri, Phys. Rev. Lett. {\bf 114}, 193602 (2015).

\bibitem{ExpHuang}
K. Huang, H. Le Jeannic, V. B. Verma, M. D. Shaw, F. Marsili, S. W. Nam, E Wu, H. Zeng, O. Morin, and J. Laurat,
Phys. Rev. A {\bf 93}, 013838 (2016).

\bibitem{nonclassty1}
A. Mari, K. Kieling, B. M. Nielsen, E. S. Polzik, and J. Eisert, Phys. Rev. Lett. {\bf 106}, 010403 (2011).

\bibitem{nonclassty2}
A. Miranowicz, M. Bartkowiak, X. Wang, Y.-x. Liu, and F. Nori, Phys. Rev. A {\bf 82}, 013824 (2010).

\bibitem{nonclassty3}
J. S. Ivan, S. Chaturvedi, E. Ercolessi, G. Marmo, G. Morandi, N. Mukunda, and R. Simon,
Phys. Rev. A {\bf 83}, 032118 (2011).

\bibitem{nonclassty4}
J. Park, J. Zhang, J. Lee, S.-W. Ji, M. Um, D. Lv, K. Kim, and H. Nha,
Phys. Rev. Lett. {\bf 114}, 190402 (2015).

\bibitem{nongaussty1}
J. S. Ivan, M. S. Kumar, and R. Simon, Quantum Inf. Process. {\bf 11}, 853 (2012).

\bibitem{nongaussty2}
M. G. Genoni, M. G. A. Paris, and K. Banaszek, Phys. Rev. A {\bf 76}, 042327 (2007);
{\it ibidem} {\bf 78}, 060303 (2008).

\bibitem{nongaussty3}
M. G. Genoni and M. G. A. Paris, Phys. Rev. A {\bf 82}, 052341 (2010).

\bibitem{nongaussty4}
M. Barbieri, N. Spagnolo, M. G. Genoni, F. Ferreyrol, R. Blandino, M. G. A. Paris, P. Grangier, and R. Tualle-Brouri,
Phys. Rev. A {\bf 82}, 063833 (2010).

\bibitem{nongaussty5}
P. Marian and T. A. Marian, Phys. Rev. A {\bf 88}, 012322 (2013).

\bibitem{nongaussty6}
C. Hughes, M. G. Genoni, T. Tufarelli, M. G. A. Paris, and M. S. Kim, Phys. Rev. A {\bf 90}, 013810 (2014).

\bibitem{SqueezBellEngineer}
F. Dell'Anno, D. Buono, G. Nocerino, A. Porzio, S. Solimeno, S. De Siena, and F. Illuminati,
Phys. Rev. A {\bf 88}, 043818 (2013).

\bibitem{AdessoIlluminati2005}
G. Adesso and F. Illuminati, Phys. Rev. Lett. {\bf 95}, 150503 (2005).

\bibitem{TelepGainBowen}
W. P. Bowen, N. Treps, B. C. Buchler, R. Schnabel, T. C. Ralph, T. Symul, and P. K. Lam, IEEE J. Sel. Top. Quant. {\bf 9}, 1519 (2003).

\bibitem{MarianCVTelep}
P. Marian and T. A. Marian, Phys. Rev. A {\bf 74}, 042306 (2006).

\bibitem{LeonhardtRealHomoMeasur}
U. Leonhardt and H. Paul, Phys. Rev. A {\bf 48}, 4598 (1993).

\bibitem{WallsMilburn}
D. Walls and G. Milburn, \textit{Quantum Optics} (Berlin, Springer, 1994).

\bibitem{DecohReview}
A. Serafini, M. G. A. Paris, F. Illuminati, and S. De Siena, J. Opt. B: Quantum Semiclass. Opt. {\bf 7}, R19 (2005).

\bibitem{Schnabel}
H. Vahlbruch, A. Khalaidovski, N. Lastzka, C. Gr\"{a}f, K. Danzmann, and R. Schnabel, Class. Quantum Grav. {\bf 27}, 084027 (2010);
C. E. Vollmer, C. Baune, A. Samblowski, T. Eberle, V. H\"{a}ndchen, J. Fiur\'{a}sek, and R. Schnabel, Phys. Rev. Lett. {\bf 112}, 073602 (2014);
H. Vahlbruch, M. Mehmet, S. Chelkowski, B. Hage, A. Franzen, N. Lastzka, S. Gossler, K. Danzmann, and R. Schnabel, Phys. Rev. Lett. {\bf 100}, 033602 (2008);
T. Eberle, S. Steinlechner, J. Bauchrowitz, V. H\"{a}ndchen, H. Vahlbruch, M. Mehmet, H. M\"{u}ller-Ebhardt, and R. Schnabel, Phys. Rev. Lett. {\bf 104}, 251102 (2010);
M. Mehmet, S. Ast, T. Eberle, S. Steinlechner, H. Vahlbruch, and R. Schnabel, Opt. Express {\bf 19}, 25764 (2011).

\bibitem{TelepZubairy}
L.-Y. Hu, Z. Liao, S. Ma, and M. S. Zubairy, Phys. Rev. A {\bf 93}, 033807 (2016).

\end{thebibliography}
\end{document}